\documentclass[12pt]{article}
\usepackage{mathtext}
\usepackage{graphicx}
\usepackage[T2A]{fontenc}
\usepackage[utf8]{inputenc}
\usepackage{amsfonts}
\usepackage{amssymb}
\usepackage{amsmath}
\usepackage{pgfplots}
\usepackage{braket}
\usepackage{tikz}
\usepackage{float}
\renewcommand{\b}{\beta}

\newcommand{\e}{\epsilon}

\newcommand{\w}{\omega}
\newcommand{\s}{\sigma}
\newcommand{\la}{\lambda}
\renewcommand{\a}{\alpha}
\begin{document}
\title{The complexity of a quantum system and the accuracy of its description}
\author{Yuri I. Ozhigov$^{1,2}$,\\
{\it 
1. Moscow State University of M.V.Lomonosov, 
} \\
{\it Faculty VMK,}
\\
{\it 2. Institute of physics and technology of K.A.Valiev}
\\ 
}
\maketitle

\begin{abstract}
The complexity of the quantum state of a multiparticle system and the maximum possible accuracy of its quantum description are connected by a relation similar to the coordinate-momentum uncertainty relation. The coefficient in this relation is equal to the maximum number of qubits whose dynamics can be adequately described by quantum theory, and therefore it can be determined experimentally through Grover search algorithm. Such a restriction of the Copenhagen formalism is relevant for complex systems; it gives a natural description of unitary dynamics together with decoherence and measurement, but also implies the existence of a minimum non-zero amplitude size, as well as a restriction on the equality of bases in the state space. The quantization of the amplitude allows us to formally introduce a certain kind of determinism into quantum evolution, which is important for complex systems.
\end{abstract}

\section{Introduction}

Our understanding of quantum theory has evolved greatly since its inception. If until about 80-90 years of the 20th century, as a rule, simple, from the classical point of view, systems were studied: individual atoms, molecules or ensembles consisting of identical particles that could be reduced to separate independent simple objects, then in recent decades the focus of research has shifted towards more complex systems. In particular, the relevance of microbiology and virology has also aroused physicists  interest in studying objects related to living things, for example, the DNA molecule, which can no longer be attributed to simple systems.

Meanwhile, quantum theory, which is the basis of our understanding of the microcosm, and, therefore, an accurate understanding of complex systems, has a very rigid and well-defined mathematical apparatus based on the matrix technique. The predictions of quantum mechanics have always proved to coincide very precisely with experiments on simple systems that are traditional for physics, but for complex systems this theory meets a fundamental obstacle. The very procedure of obtaining theoretical predictions requires such unimaginable computational resources that we will never have them at our disposal.

If for simple systems the procedure of computation the quantum state had no relation to the physics of its evolution and was only a technical problem, then for complex systems the situation is different. Here, the computation process is the main part of the definition of the quantum state itself, and therefore should be considered as a physical process, and the device that implements this computation is an integral part of any experiment with complex systems at the quantum level.

This computing device is an abstract computer that simulates the evolution of the complex system under consideration. Thus, all restrictions on this computer, following from the theory of algorithms, have the status of physical laws; and these laws have absolute priority over physical laws in the usual sense in the case of complex systems and processes.

This is a new situation that did not exist in classical physics, where the procedure for obtaining theoretical predictions was not very complicated. In any case, the complexity there has almost always been within the reach of classical supercomputers, which are created mainly to cover processes from the point of view of classical physics - by the number of particles in the system under consideration. In quantum mechanics, the complexity increases exponentially with the number of particles, and the classical way of computing becomes unacceptable. This was strictly proved by the discovery of theoretically possible (from the point of view of the standard - Copenhagen quantum theory) processes that cannot be modeled on any classical supercomputers - the so-called fast quantum algorithms (\cite{Sh}, \cite{Gr}). 

The attempt to circumvent the complexity barrier with the help of a quantum computer proposed by R. Feynman (\cite{Fey}) has given us a lot to understand the microcosm and some interesting applications, for example, in cryptography and metrology. However, this attempt did not solve the main problem: the scaling of a fully functional quantum computer is very questionable due to decoherence. Decoherence occurs as a result of spontaneous measurements of the states of the simulated system from the environment, which is traditionally considered in the framework of the concept of an open quantum system in contact with the environment (see \cite{BP}), so that the influence of the environment is reduced to uncontrolled measurements of the state of the original system.

Thus, decoherence is a fundamental factor that cannot be eliminated with the help of mathematical techniques, such as error correction codes (they begin to really work only for a quantum computer with more than a hundred qubits). If we set the task of modeling complex systems at the quantum level, decoherence should be embedded in the quantum formalism itself, and not introduced into it as an extraneous influence. The deviation from the linear unitary law of evolution resulting from decoherence must be naturally justified mathematically.

There is a complexity barrier to matrix formalism. 

The more precisely we know the amplitudes of the quantum state, the simpler it should be. If the state is complex, we will not be able to determine its amplitudes exactly. Given a system of $n$ particles - qubits, to apply quantum mechanics we try to learn their state as precise as possible. 

A quantum state $|\Psi\rangle$ cannot be the state of a single system of qubits; the wave vector is a characteristic of a huge number of equally prepared such systems. Thus, the quantum state $| \Psi\rangle $ characterizes a certain imaginary apparatus that produces exactly the same ensembles consisting of $n$ qubits.

The accuracy of the quantum state is the accuracy of determining its amplitudes of the basic states using measurements: the more copies of the state we have, the more accurately we can determine these amplitudes by simultaneously measuring all these copies.

Let there be a system of $n$ qubits about which we think that it is in the quantum state $|\Psi\rangle$.  We will call the accuracy of this state the maximum possible number of such equally prepared ensembles. The accuracy of a state is thus the maximum possible number $A$ of copies of this state that can be available to us at the same time, when we can measure them. 

We arrange such $n$ qubit ensembles $S_j$ having the same states in the form of a sequence of $nA$ qubits of the form
$$
S_1,S_2,...,S_A,\ \ S_j=(s_1^j,s_2^j,...,s_n^j),
$$
thus obtaining the memory of some anstract Main Computer (MC), from which we can learn what this state is. The memory of MC cannot be unlimited, hence there is the constant  $Q$, such that  
\begin{equation}
\label{1}
An\leq Q.
\end{equation}

But the number $n$ of qubits cannot be the exact measure of the complexity of the state $|\Psi\rangle$, even if these qubits are all entangled. To define the real complexity we have to use the so called canonical transformation, which can radically reduce the number $n$ of qubits without changing the state $|\Psi\rangle$ substantionally. 

For the right definition of complexity of a state $|\Psi\rangle$ the following inequality takes place:

\begin{equation}
\label{und}
AC\leq Q.
\end{equation} 

We will argue in favor of such a ratio of complexity and accuracy in the general case, for complex particles; in particular, we will show that the quantization of the amplitude makes it possible to introduce a certain determinism into the quantum formalism, the nature of which is not reduced to the classical one.

\section{The Main Computer}

A complex system is a system whose behavior cannot be reduced to independent particles. Predictions of the behavior of such a system can be obtained only by relying on quantum mechanics and computer ideology, since analytical techniques do not work here. Therefore, we must introduce the concept of a Main Computer (MC), which adequately represents real processes for us - an abstract computing device, the laws of which will be for complex systems (and only for them!) have priority over physical laws, the scope of which is unconditionally limited to simple systems and processes.

A physical prototypes of a Main Computer can have only limited power; such devices will be able to adequately represent the processes traditionally related to chemistry, as well as to those areas of physics in which quantum methods are successfully applied, for example, to electrodynamics. Nuclear physics does not yet belong to such processes, and its complexity radically exceeds the electrodynamics  complexity (see \cite{Oz}).

Thus, we have, within the framework of this limitation of the computational capabilities of the MC, a balance between the complexity and accuracy of the representation of the state vector, which must be found specifically in each specific case.

\begin{figure}
\centering
\includegraphics[scale=0.57]{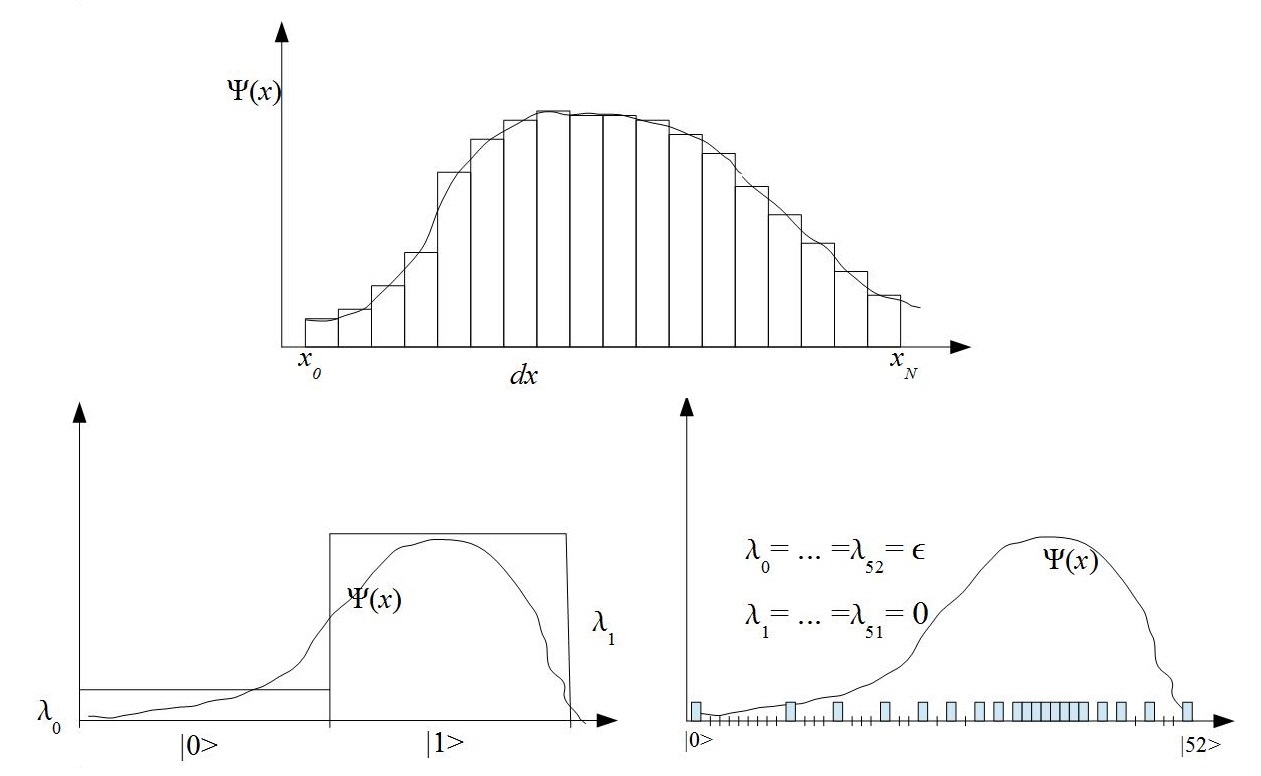}
\caption{Representation of the state vector. Curves represent the hypothetic wave function $|\Psi\rangle$, predicted by Copenhagen quantum theory. Rectangles denote the information about it, which we can obtain via MC. Bottom left - the main computing resource captures accuracy: $|\Psi\rangle=\lambda_0|0\rangle+\lambda_1\rangle $; idealized case where we reduce the number of basic states to 2, as for a particle in two hole potential - this gives the satisfactory agreement with experiment.  The top - the resource is divided equally between accuracy and complexity. The best agreement with experiments; there is the typical area of application of quantum mechanics. Bottom right - the main resource is captured by complexity: $|\Psi\rangle=\sum\limits_{j=0}^{N-1}\varepsilon |j\rangle,\ \varepsilon \in\{ 0,\epsilon\}$. Our knowledge is limited by only one basic state, which we extract from the single measurement. Here we have to follow the trajectory of one quantum of amplitude (see below). }
\label{fig:discr1}
\end{figure}

For one qubit we can find its amlitudes with the most possible precision. For the simple systems with low complexity, which was in the focus of attention of the physics of 20th century, the possible precision was equal to the accuracy of experimental results. For such systems with  of intermediate complexity we were able to determine the amplitudes more accurately. For the more complex systems, as the prototypes of quantum computer, we already meet difficulties that is called the decoherence.
For an extremally complex system $A=1$ we have only one sample of it; we can make only one measurement, which means that we can receive only one basic state.  The figure \ref{fig:discr1} represents these cases.

\section{Complexity of the quantum state}

In order to give a correct definition of the complexity of the quantum state of a system of $ n $ particles, we must first consider the canonical transformation - the main method of reducing such complexity.

Any coordinate on a unit segment is a real number represented as its binary expansion $2^{-l}\sum\limits_{j=0}^{l-1}a_j2^j$ with an accuracy of $2^{-l}$, where $a_j=0.1$ are the values of $l$ qubits representing this coordinate\footnote{To represent a coordinate on any other segment, we need to apply a suitable linear transformation, for example, on the segment $[-2^{l/2}, 2^{l/2}]$ the approximate qubit representation looks like this: $2^{-l/2} \sum\limits_{j=0}^{l-1}a_j2^j-2^{l/2-1}$.} By ordering the qubit values lexicographically, we get a standard ordering of basis vectors, in which any operator will have a certain matrix.

Let the classical state of the particle $i$ be the real vector $x_i$. Then the classical state of the system of $n$ particles will be a vector of the form $ \bar x=(x_1, x_2,..., x_n)$. Let $\bar x'$ and $\bar x"$ be two vectors with non-empty sets of coordinates such that their Cartesian product coincides with $\bar x$. This means that we have split the set of particles into two non-empty subsets $X'$ and $X"$, so that the given vectors are the sets of coordinates of these subsets. Let $H (\bar x)$ be the Hamiltonian of this system, having the form

\begin{equation}
\label{ham}
H(\bar x)=H_1(\bar x')+H_2(\bar x'');
\end{equation}
here, we assume by default that $H(\bar x')$ is $H(\bar x')\otimes I (\bar x")$, that is, on particles not included in the first subset, this term of the Hamiltonian acts as an identical operator, and similarly with the second term. Then we call the Hamiltonian $H$ reducible. Let $X'$ be the maximum subset of the components of the vector $\bar x$ in terms of the number of elements, such that the equality \eqref{ham} holds, and the Hamiltonian $H_1(\bar x')$ is not reducible. Then $X'$ is called the kernel of the given Hamiltonian.

We will assume by default that any quantum evolution begins with the basic state of the system under consideration. Since \eqref{ham} implies the equality $exp(-\frac{i}{\hbar}H)=exp(-\frac{i}{\hbar})H_1\otimes exp (- \frac{i}{\hbar}) H_2$, we see that the kernel of the Hamiltonian is the maximum set of particles whose states in quantum evolution with the Hamiltonian $H$ can be entangled; we denote the number of particles in this set by $ \nu(H)$, and call it the {\it naive} complexity of this Hamiltonian.

Consider the transformation of the coordinates of the particles of the form

\begin{equation}
\label{trans}
q_i=q_i(x_1,x_2,...,x_n),\ i=1,2,...,n;
\end{equation}
denote $\bar q =(q_1, q_2,..., q_n)$ and let $H_q=H(x_1(\bar q),x_2(\bar q),..., x_n(\bar q))$ is the original Hamiltonian written in terms of the new variables $q_i$, $\bar x=\bar x(\bar q)$ is the inverse of \eqref{trans}. We will introduce virtual particles with coordinates $q_1, q_2,..., q_n$, which we will call quasiparticles.

The classical state of the system in the initial representation is a set of specific values $x_1,x_2,..., x_n$. Then each classical state will correspond to the classical state of the same system, obtained using the formulas \eqref{trans}. The basis vector of a Hilbert space passes into another basis vector of the same space. The standard ordering of the basis vectors, according to the qubit representation of coordinates, will pass to another ordering, that is, the coordinate transformation \eqref{trans} is a permutation of the basis vectors of the Hilbert space of quantum states.

In this case, the qubits representing the coordinate values will already have a different, new meaning. In the new coordinates, for quasiparticles, the Hamiltonian will have a different form $H_q$. We will call the coordinate transformation \eqref{trans} canonical if $ \nu (H_q)$ is minimal. In this case, the transition to quasiparticles will mean a reduction in the complexity of the original Hamiltonian. So, the canonical transformation is a permutation of the bisis vectors that minimizes the complexity of the Hamiltonian.

The transition to the description of evolution in the form of quasiparticles has the form $H=\tau^{-1} H_q \tau$ where the permutation of basis vectors $\tau$ is the transformation of the transition to quasiparticles under the canonical transformation. Then the representation of the evolution operator is $exp (- \frac{i}{\hbar}Ht)=\tau^{-1}exp(-\frac{i}{\hbar}H_qt)\tau$ requires less computational resources than the direct calculation of $exp (- \frac{i}{\hbar}Ht)$, since the main resource is spent on the kernel, which for a quasi-partial representation will have a minimum size.

An example is a chain of interacting harmonic oscillators, for which the Fourier transform of their coordinates $x_i$ is canonical, and the kernel of the quasiparticle representation-in terms of phonons - will generally consist of a single particle, that is, the phonons are all independent. Here we have the maximum reduction of the kernel. 

A simpler example would be the Hamiltonian $H$ of a closed chain of 4 interacting qubits, which is reduced by the canonical transformation $CNOT$ to a completely reduced Hamiltonian of the form
 $H_q=\s_x^{(1)}\otimes I_2+I_1\s_x^{(2)}$: 
$$
H=
\begin{pmatrix}
&0&1&0&1\\
&1&0&1&0\\
&0&1&0&1\\
&1&0&1&0
\end{pmatrix}
=CNOT
\begin{pmatrix}
&0&1&1&0\\
&1&0&0&1\\
&1&0&0&1\\
&0&1&1&0
\end{pmatrix}
CNOT.
$$
Note that the permutation of the basis vectors, which is a canonical transformation, must be an entangling and simultaneously disentangling operator in the state space, since it reduces the kernel of the Hamiltonian. So, for example, the $CNOT $ operator applied repeatedly to the state $|00...0\rangle+|11...1\rangle$ completely untangles this state.

The opposite example is given by the Tavis-Cummings Hamiltonian for $n$ two-level atoms interacting with the resonant mode field in the optical cavity. Here there is a basic state of the field and atoms of the form $ |n \rangle_{ph}|00...0 \rangle_{at}$, such that the column of the Hamiltonian matrix corresponding to this state consists of the numbers $g\sqrt{n}$ and one number $n\hbar\w$, and such a column is the only one. No Hamiltonian of the form \eqref{ham} can have this property even at $n=2$, so there is no non-identical canonical transformation for the Tavis-Cummings Hamiltonian.

The quantum complexity of the Hamiltonian $H$, denoted by $C(H)$, is the minimal naive complexity of the operators $ \tau^{-1} H \tau$ over all possible permutations of $\tau $ basis vectors. For the above examples, the complexity of the Hamiltonian is 1, that is, it can be completely reduced by the canonical transformation.

The complexity  of the quantum state $|\Psi\rangle$ is determined in a similar way. Its {\it naive} complexity is $\nu(|\Psi\rangle)$ is defined as the number of particles in the maximum tensor divisor $|\Psi_1\rangle$ of the state $|\Psi\rangle=|\Psi_1\rangle\otimes |\Psi_2\rangle$.

The quantum complexity $C(|\Psi\rangle)$ of the state $| \Psi\rangle$ is the minimal naive complexity of the state $\tau |\Psi\rangle$ over all permutations $\tau$ of the basis states. 

For example, the quantum complexity of the generalized GHZ state $ |GHZ\rangle=\frac{1}{\sqrt 2}(|00...0\rangle+|11...1\rangle)$ is equal to 1, since it can be untangled by successive CNOT operators. If we start from the basis state $|\Psi(0)\rangle$ in the canonical representation of the Hamiltonian $ H $, then an evolution with $H$ will only contain states with a complexity of no more than $C(H)$. 

We can now formulate a hypothesis about the relation of the form "accuracy - complexity" in the final form:
\begin{equation}
C(|\Psi\rangle) A(|\Psi\rangle) \leq Q,
\label{main}
\end{equation}
where $Q$ is the maximum number of completely entangled qubits that cannot be disentangled by any permutation of the basis states.

Consider, as an example, the state of a set of $n$ qubits of the form 
\begin{equation}
\label{GSAstate}
|\Psi_{GSA}(t)\rangle= \a\sum\limits_{j\neq j_0,0\leq j<N}|j\rangle+\b|j\rangle,
\end{equation}
where $\a=cos(t)/\sqrt{N-1},\ \b=sin(t)$ for some $t$, and $N=2^n$. The quantum complexity of this state is $ n$ if $t\neq k\pi/2$ for no integer $k$. Indeed, this superposition has the property that all its basic components, except for exactly one, have the same non-zero amplitude, and one has a different amplitude from them.

This property is preserved under any permutation of the basis states, that is, under any quasiparticle representation. But if the state is reducible, then it should have the form $\la_1|i_1\rangle+\la_2|j_2\rangle+...)\otimes(\la_3|j_3\rangle+\la_4|j_4\rangle+...)$ for some basis $|j_i\rangle$, and such a superposition cannot contain exactly 2 amplitude values for any basis states, since there must either be at least 3 different non-zero amplitude values of the components, or it must have only two different non-zero amplitude values that correspond to two groups of basis states containing an equal number of terms. Both of these possibilities are excluded for states of the form \eqref{GSAstate}.

\section{Amplitude granularity}

Here we show the possible way how to prolong quantum formalism over  the border $Q$. This is the kind of quantum determinism that is not reducible to the known quasiclassical effects.

The qubit representation of the classical coordinates and impulses determines the grain size of the amplitudes. For any expansion of the quantum state $|\Psi\rangle$ over an arbitrary orthonormal set of basis states $|\psi_j\rangle$ of the form 
\begin{equation}
\label{state}
|\Psi\rangle = \sum\limits_{j\in J}\la_j|\psi_j\rangle,\ \la_j\neq 0,
\end{equation}
the amplitudes of $\la_j$ must be bounded from below modulo some nonzero constant.

To preserve the principle of linearity in the region where the nonzero value of this constant does not play a role, we must assume that any amplitude has the form 
\begin{equation}
\label{quant}
\la_j = \e n_j+i\e m_j
\end{equation}
where $n_j,m_j\in Z$ are integers, $\e>0$ is a constant that is the quantum of the amplitude. This restriction of the matrix formalism entails the rejection of the absolute equivalence of any bases in the space of quantum states, which also manifests itself only at sufficiently small amplitudes $ \la_j$, and, accordingly, large sets of coherent states $J$.

However, modeling does not provide for such equivalence of bases; it is only an algebraic technique that cannot be experimentally tested for complex systems.
The smallest modulo possible nonzero amplitude is thus $\e$. This restriction is very well consistent with the probabilistic nature of the state vector, since to determine the value of $|\la_j|^2$ with an accuracy $ \delta$, it is necessary to conduct about $1/\delta$ measurements of equally prepared samples of the original system; for complex systems with small amplitudes $\la_j$, this will be possible only if the minimum probability $\e^2$ is separated from zero to obtain the most unlikely outcome.

We can determine the value of $ \e$ by "smearing" the amplitude over as many basis classical states of the system as possible, for which the presence in them can simply be detected by measurement. If all the amplitudes are equal to $\la_j=\e$, we get for the total number of coherent basis states the estimate $|J|=1/\e^2$ - as the inverse square of the amplitude quantum; it equals to the maximal dimension $2^Q$ of Hilbert space for quantum states, in which we can use quantum formalism. In the \eqref{state} expansion, the amplitudes of $\la_j$ have no physical dimension, the dimension has the basis states of $|\psi_j\rangle$. The number of qubits whose possible quantum states are physically realizable is equal to the constant $Q$, for which we get the expression $\e=2^{-Q/2}$.

Note that the discrete representation of amplitudes in the form of \eqref{quant} allows us to naturally include state measurements in its unitary evolution. The hard contact of the system in the state $|\Psi\rangle=\sum\limits_{j\in J}\la_j|j\rangle$ with the measuring device means the inclusion of the states of this device in the system, that is, the transition to the state of the extended system of the form $|\Psi_{ex}\rangle=\sum\limits_{j\in J,\nu_j\in{\cal N}_j}\mu_{j,\nu_j}|j,\nu_j\rangle$, where, under the condition of the applicability of quantum mechanics, all $\mu_{j,\nu_j}$ must be minimal, which means that they equal $\epsilon$. Since contact with the environment is a unitary evolution, we have $|\la_j|^2=\sum\limits_ {\nu_j\in{\cal N}_j}|\mu_{j,\nu_j}|^2$ that is, the measurement will be an arbitrary choice of one of the states $\nu_j$ of the measuring device and we get the standard urn scheme from probability theory.

\section{Experimental finding of a constant $Q$}

We can find the approximate value of $Q$ by the building the state of the form \eqref{GSAstate}. These states have the memory of a quantum computer when implementing the Grover search algorithm (GSA) with a single target state $|j_0 \rangle$ (see \cite{Gr}). Let $N=2^n$. Let's put $t_0=arcsin(1/\sqrt{N})$, $|\Psi(t_0)\rangle=\frac{1}{\sqrt N}\sum\limits_{j=0}^{N-1}|j\rangle=(\frac{1}{\sqrt 2}(|0\rangle+|1\rangle)^{\otimes n}$; the complexity of this state is 1. This is the initial state for the GSA algorithm. As soon as the first step of the algorithm is performed, $t$ becomes equal to $t_0+2t_0$ and we get a state of complexity $n$, of the form \eqref{GSAstate}. Already at the first step of the algorithm, the complexity jumps from one to the maximum value  $n$ for $n$ qubit particles. At the first step of the GSA algorithm, we thus will go beyond the limits of acceptable states with ampitudes of the form \eqref{quant}, which have the property \eqref{main}.

Thus, we can estimate the constant $Q$ from above, increasing the total number $n$ of qubits to the limit when the GSA stops working correctly. Here, by correct operation, we mean the possibility to raise the amplitude of the target state by an amount of the order of $1/ \sqrt{2}$ compared to the others, which can be fixed by quantum tomography, since the amplitudes of other states will have the order of $1/\sqrt N$. For a more rough estimate, fixing the jump in the amplitude by $1/\sqrt N$ compared to the main mass is also suitable, but this is only possible for small $n$, not exceeding 20.

The question of what will happen to the real state if the amplitudes calculated according to standard quantum theory become less than $\e$ is formally open. However, it is natural to assume that small amplitudes should simply disappear, with a corresponding renormalization of the remaining state. This means that implementing GSA near the boundary $N\approx 2^Q$ of the dimensionality we will get the target state very quickly, much faster than when implementing GSA in a normal model. However, this will only happen with an ideal implementation of GSA; in practice, the amplitudes of the main mass of states in $|\Psi_{GSA}(t)\rangle$ in \eqref{GSAstate} cannot be exactly the same, so that zeroing will not occur simultaneously for all states, which can greatly distort the picture.

\section{Equilibrium states}

We will begin by describing the classes of states for which the amplitude quantization is introduced in the most visual way.

For a complex number $z=a+ib,\ a,b\in R$, we introduce the notation $\{ z\}=|a|+|b|$.
For the quantum state $|\psi\rangle$, we define $\{ \psi\}=\sum\limits_{i=0}^{N-1}\{\langle i|\psi\rangle\}$.

Let $A$ be a linear operator, and $|j \rangle$ be some basis vector, $j\in\{ 0,1,..., N-1\}$. Define $|a_j\rangle=A|j\rangle$. We call the state vector $| \Psi\rangle$ equilibrium with respect to the operator $A$, if all the numbers $\{a_j\}$ are the same for all the basis components $|j\rangle$ that are included in it with nonzero amplitudes. We note that this definition depends on the basis of the space.

As an example, consider the Hamiltonian of a one-dimensional particle moving in the potential $V$ in the coordinate basis: $H=\frac{p^2}{2m}+V$. We will reduce the matrix of this Hamiltonian, assuming that there are no too long transitions of a given particle in space. Then the equilibrium states in the coordinate basis for this Hamiltonian will be exactly the states $| \Psi\rangle$, all the basis components of which have the same potential.

An important class of multiparticle equilibrium states are connected states. Here is an example of such a state. 
Consider $k$ of two-level atoms in an optical cavity holding photons with the transition energy between the ground and excited levels of the atoms. We choose a basis consisting of vectors of the form $|n\rangle_{ph}|m_1,m_2,..., m_k\rangle_{at}$, where $n$ is the number of photons in the cavity, $m_j\in\{ 0,1\rangle$ is the state of the atom $j$, ground and excited. Let $g_j,\ j=1,2,..., k$ be the forces of the interaction of atoms with the field. Then the dynamics of the system of atoms and the field under the condition $g_{j}/\hbar\omega\ll 1$, where $ \omega$ is the frequency of the cavity, will obey the Schredinger equation with the Tavis-Cummings Hamiltonian in the RWA approximation:
\begin{equation}
H_{TC}^{RWA}=\hbar\omega (a^+a+\sum\limits_{j=1}^k\sigma^+_j\sigma_j)+a^+\bar\sigma+a\bar\sigma^+,\ \ \bar\sigma=\sum\limits_{j=1}^kg_j\sigma_j,
\label{exampleHam}
\end{equation}
where $a, a^+$ are the standard field operators of photon annihilation and creation, and $\sigma_{j},\sigma_j^+$ are the atomic relaxation and excitation operators of the atom $j$. The connected states in such a system will be for $k=2$ only for $g_1=g_2$, and this will be either one of the basis states, or the states $|n\rangle_{ph} (\a|10\rangle_{at}+ \b|01\rangle_{at})$, of which in all but the singlet state $\b=-\a$ the atoms will interact with the field. All such states will be  equilibrium.

The general definition of connectivity looks like this.

Let $H$ be a Hamiltonian in the state space of $n$ qubits. If a qubit is associated with a real or virtual two-level particle, $H$ can be, for example, the Tavis-Cummings Hamiltonian or some modification of it. Let $S_n$ be a group of permutations of qubits that are naturally extended to operators on the entire space of quantum states ${\cal H}$, namely: on the basis states, the permutation $\eta\in S_n$ acts directly, and $\eta\sum\limits_j|j\rangle=\sum\limits_j\eta|j\rangle$.

Denote by $G_H$ the subgroup $S_n$ consisting of all permutations of qubits $ \tau$ such that $[H,\tau]=0$. Let $A\subseteq\{ 0,1,..., 2^n-1\}$ be a subset of the basis states of the $n$ - qubit system. Its linear shell $L (A)$ is called a connected subspace with respect to $H$ if for any two states $|i \rangle,\  |j\rangle\in A$ there exists a permutation of qubits $\tau\in G_H$ such that $\tau (i)=j$. A state $|\Psi\rangle$ of $n$ - qubit system is called connected with respect to $H$ if it belongs to a connected subspace with respect to $H$, and $H|\Psi\rangle\neq 0$.

The connectedness of a state means that all its nonzero components are obtained from one another by permutations of those particles that behave in the same way with respect to a given Hamiltonian. 
The above example of the state $ |n\rangle_{ph}(\a|10\rangle_{at}+ \b|01\rangle_{at})$ will obviously be connected, since the permutation of atoms interacting equally with the field does not change the Hamiltonian. The states of the form $|n\rangle_{ph}(\a|10\rangle_{at}+ \b|01\rangle_{at}+c|00\rangle_{at}+d|11\rangle_{at})$, for non-zero values of the amplitudes $a,b,c, d$ will not be connected. 
\bigskip

{\it Lemma.

If $|\Psi\rangle=\sum\limits_j\la_j|j\rangle$ is connected with respect to $H$, then any two columns of the matrix $H$ with numbers $j_1,\ j_2$, such that $\la_{j_1}$ and $\la_{j_2}$ are nonzero, differ from each other only by the permutation of elements. The same is true for the unitary evolution matrix $U_t=exp(-\frac{i}{\hbar}Ht)$. }
\bigskip

Indeed, for such basis states $j_1$ and $j_2$, according to the definition of $H$ - connectivity, there exists $\tau\in G_H$ such that $j_2=\tau(j_1)$. The columns numbered $j_1,\ j_2$ consist of the amplitudes of the states $H|j_1\rangle$ and $H|j_2\rangle$, respectively. From the commutation condition, we have $\tau H|j_1\rangle=H\tau |j_1\rangle=H|j_2\rangle$, and this just means that the column $j_2$ is obtained from the column $j_1$ by the permutation of elements induced by $\tau$. Moving on to the evolution matrix $U_t$, we see that the switching relation $\tau U_t|j_1\rangle=U_t\tau |j_1\rangle=U_t|j_2\rangle$ will hold for it as well, which is what is required. Lemma is proven.

It follows from Lemma that the states connected with respect to the Hamiltonian $H$ are equilibrium with respect to $H$ and with respect to the evolution operator $U_t=e^{-\frac{i}{\hbar}Ht}$ corresponding to this Hamiltonian.

\section{Amplitude quanta and determinism\label{quantampl}}

The exact description of the dynamics, even in the classical framework, has some degree of nondeterminism or stochasticity (see \cite{det1},\cite{det2}). The advantage of the quantum language lies in the precise limitation of this stochasticity to the state vector, which, according to the main thesis of Copenhagen mechanics, provides an exhaustive description of the dynamics of microparticles.

The possibility of introducing determinism into quantum theory, which interested researchers at the beginning of the history of quantum physics, has not lost the interest of researchers (see \cite{det3}) and is again becoming relevant for complex systems; for example, for systems of extreme complexity, the DNA molecule determines the trajectory of its owner with an accuracy unattainable in physical experiments. This higher type of determinism must have an analog for the simple systems described by standard quantum theory.

We describe a possible specific form of determinism at the level of amplitude quanta.

Our goal is to show that if the state $|\Psi\rangle$ is equilibrium with respect to the evolution operator $U_t$, then the amplitudes of all the basic states in $|\Psi\rangle$ can be divided into small portions - quanta of amplitude, so that for each quantum its trajectory will be uniquely determined when $U_t$ acts on a previously fixed time interval $t$, in particular, it will be uniquely determined also, which other quantum of amplitude it will contract with when summing the amplitudes to obtain the subsequent states.

This fact is also true for the arbitrary operator $A$, for which we will formulate the quantization of the amplitude. 

Let $| \Psi\rangle$ be an arbitrary equilibrium state with respect to $A$, the decomposition of which in terms of the basis has the form

\begin{equation}
\label{connected_state}
|\Psi\rangle=\sum\limits_j\la_j|j\rangle.
\end{equation}

We introduce the important concept of the amplitude quantum as a simple formalization of the transformation of a small portion of the amplitude between different basis states when multiplying the state vector by the matrix $A$. Let $T=\{ +1,-1,+i,-i\}$ be a set of 4 elements, which are called amplitude types: real positive, real negative, and similar imaginary ones. The product of types is defined in a natural way: as a product of numbers. The quantum of the amplitude of the size $\varepsilon>0$ is called a list of the form
\begin{equation}
\label{quanta}
\kappa=(\varepsilon,id, |b_{in}\rangle, |b_{fin}\rangle, t_{in},t_{fin}),
\end{equation}
where $|b_{in}\rangle,\ |b_{fin}\rangle$ are two different basis states of the system, $id$ is a unique identification number that distinguishes this quantum from all others, $t_{in}, t_{fin}\in T$. Transition of the form $|b_{in}\rangle\rightarrow\ |b_{fin}\rangle$ is called a state transition, $t_{in}\rightarrow t_{fin}$ is called a type transition. Let's choose the identification numbers so that if they match, all the other attributes of the quantum also match, that is, the identification number uniquely determines the quantum of the amplitude. In this case, there must be an infinite number of quanta with any set of attributes, except for the identification number. Thus, we will identify the quantum of the amplitude with its identification number, without specifying this in the future. Let's introduce the notation:
$$
t_{in}(\kappa)=t_{in},\ t_{fin}(\kappa)=t_{fin},\ s_{in}(\kappa)=b_{in},\  s_{fin}(\kappa)=b_{fin}.
$$

The transitions of states and types of amplitude quanta actually indicate how a given state should change over time, and their choice depends on the choice of $A$; the size of the amplitude quantum indicates the accuracy of a discrete approximation of the action of this operator using amplitude quanta.


The set of $ \theta $ quanta of the amplitude of the size $\varepsilon$ is called the quantization of the amplitude of this size, if the following condition is met:

{\bf Q}. In the set $\theta$, there are no such quanta of the amplitude $\kappa_1$ and $\kappa_2$ that their state transitions are the same, $t_{in}(\kappa_1)=t_{in}(\kappa_2)$ and $t_{fin}(\kappa_1)=-t_{fin}(\kappa_2)$, and there are no such quanta of amplitude $\kappa_1$ and $\kappa_2$ that $s_{in}(\kappa_1)=s_{in}(\kappa_2)$ and $t_{in}(\kappa_1)=-t_{in}(\kappa_2)$.

\bigskip

The condition {\bf Q} means that during the transition described by the symbol " $ \rightarrow$", the total value of the amplitude quantum cannot contract with the total value of a similar amplitude quantum, and also that the amplitude quanta do not contract with each other directly in the initial state record.

Quantization of the $\theta $ amplitude sets a pair of quantum states
\begin{equation}
\label{theta}
|\theta_{in}\rangle=\sum\limits_j\la_j|j\rangle, \ |\theta_{fin}\rangle=\sum\limits_i\mu_i|i\rangle, 
\end{equation}
according to the natural rule: for any basis states $|j\rangle, \ |i\rangle$, the equalities must be satisfied
 \begin{equation}
\la_j=\langle j|\theta_{in}\rangle=\varepsilon\sum\limits_{\kappa\in\theta:\ s_{in}(\kappa)=j}t_{in}(\kappa),\ \ \ \ 
\mu_i=\langle i|\theta_{fin}\rangle=\tilde\varepsilon\sum\limits_{\kappa\in\theta:\ s_{fin}(\kappa)=i}t_{fin}(\kappa),
\label{shift_}
\end{equation}
where $ \tilde\varepsilon$ is some normalization coefficient, so that the state $|\theta_{fin}\rangle$ has a unit norm, and $|\theta_{in}\rangle$ has an arbitrary nonzero one. The coefficient $\tilde\varepsilon$ does not have to coincide with $\varepsilon$, because when quantizing the amplitude, the usual norm of the state vector, in general, is not preserved; if we took $\tilde\varepsilon=\varepsilon$, then the value $\{|\Psi\rangle\}$ could only decrease at the transition $|\theta_{in}\rangle\rightarrow |\theta_{fin}\rangle$; this is exactly due to the fact that some quanta of the amplitude "contract" with each other in the second sum from the formula \eqref{shift_}.

Let's fix the dimension $dim({\cal H})$ of the state space, and we will make estimates (from above) of the considered positive quantities: the time and the size of the amplitude quantum up to an order of magnitude, considering all constants to depend only on the independent constants: $dim({\cal H})$ and from the minimum and maximum absolute values of the elements of the matrix $A$. At the same time, the term ''strict order'' will mean the evaluation of both the top and bottom positive numbers that depend only on independent constants.

Given the quantization of amplitude $\theta$ and the numbers $i, j$ of the basis states we will denote by $n_{i,j}(\theta)$ the number of elements of the set ${\cal N}_{i,j}(\theta)=\{ \kappa\in\theta:\ s_{in}(\kappa)=j,\ s_{fin}(\kappa)=i\}$. 

Let $\theta (\varepsilon)$ be some function that maps some sequence of positive numbers - values of $ \varepsilon $ converging to zero, into quantizations of the amplitude of the size $\varepsilon$. This function will be called parametric quantization of the amplitude.

The parametric quantization of the amplitude $\theta (\varepsilon)$ is called consistent with the operator $A$ and state $|\Psi\rangle$ if for some scalar functions $c(\varepsilon)$
\begin{equation}
\theta_{in}(\varepsilon)\rightarrow |\Psi\rangle,\ c(\varepsilon)\theta_{fin}(\varepsilon)\rightarrow A|\Psi\rangle \ (\varepsilon\rightarrow 0).
\label{agreement}
\end{equation}

If $A$ is the evolution operator $U_t$, then having a parametric quantization of the amplitudes $\theta(\varepsilon)$ consistent with $A$ is a completely non-trivial property of the quantum states $|\theta_{in}(\varepsilon)\rangle$, which says that it is possible to introduce a hidden parameter corresponding to the dynamics set by the evolution matrix $U_t$, and making the quantum evolution $U_t$ deterministic. Such a parameter is the quantum of the amplitude $\kappa\in\theta (\varepsilon)$, where the accuracy of the deterministic description is determined by the value $\varepsilon$.

\bigskip

{\it The amplitude quantization theorem. 

Let $A$ be an arbitrary matrix. For every equilibrium state $|\Psi\rangle$ with respect to $A$, there exists a parametric quantization of the amplitudes $\theta (\varepsilon)$, consistent with the operator $A$.

\it Proof. } 

Let we be given the equilibrium state with respect to $A$ of the form $ | \Psi\rangle=\sum\limits_j\la_j|j\rangle$ and the number $\varepsilon>0$. For $|j\rangle$ with nonzero $\la_j\neq 0$ let
\begin{equation}
\la_j=\langle j|\Psi\rangle\approx sign_{re}( \underbrace{\varepsilon+\varepsilon+\ldots +\varepsilon}_{M_j})+sign_{im}i( \underbrace{\varepsilon+\varepsilon+\ldots +\varepsilon}_{N_j} ),
\label{quanta_expansion}
\end{equation}
where $sign_{re}\varepsilon M_j+sign_{im}i\varepsilon N_j\approx \la_j$ is the best approximation of the amplitude of $\la_j$ with the accuracy of $\varepsilon$; $M_j,\ N_j$ are natural numbers, $sign_{re\ (im)}=\pm 1$. Thus, the first convergence relation from \eqref{agreement} will be satisfied, and the second relation must be satisfied if the parametric quantization is consistent with the Hamiltonian.

Let us approximate each element of the evolution matrix in the same way as we approximate the amplitudes of the initial state:
\begin{equation}
\label{hamapp}
\langle i|A|j\rangle\approx \pm(\underbrace{\varepsilon+\varepsilon+...+\varepsilon}_{R_{i,j}})\pm i (\underbrace{\varepsilon+\varepsilon+...+\varepsilon}_{I_{i,j}}),
\end{equation}
where $R_{i,j},\ I_{i, j}$ are natural numbers; the real and imaginary parts are exactly $\varepsilon$ each, and the signs before the real and imaginary parts are chosen based on the fact that this approximation should be as accurate as possible for the selected $\varepsilon$.

The amplitudes of the resulting state $A|\Psi\rangle$ are obtained by multiplying all possible expressions \eqref{quanta_expansion} by all possible expressions \eqref{hamapp}:

\begin{equation}
\label{mult}
\la_j\langle i|A|j\rangle\approx (sign_{re}M_{j}\varepsilon+i\ sign_{im}N_{j}\varepsilon )(\pm R_{i,j}\varepsilon\pm i\ I_{i,j}\varepsilon).
\end{equation}

We will expand the brackets in the right part of the expression \eqref{mult}, but we will not make abbreviations. Each occurrence of the expression $\varepsilon^2$ in the amplitudes of the resulting state after opening the brackets in the right part of \eqref{mult} will be obtained by multiplying a certain occurrence of $\varepsilon$ in the right part of \eqref{quanta_expansion} by a certain occurrence of $\varepsilon$ in the right part of \eqref{hamapp}. The problem is that the same occurrence of $\varepsilon$ in \eqref{quanta_expansion} corresponds to not one, but several occurrences of $\varepsilon^2$ in the result, and therefore we can not match the amplitude quanta directly to the occurrences of $\varepsilon$ in \eqref{quanta_expansion}.

How many occurrences of $\varepsilon^2$ in the amplitudes of the state $A|\Psi\rangle$ from the result of opening the brackets in \eqref{mult} correspond to one occurrence of $\varepsilon$ in the approximation of the amplitude $\la_j=\langle j|\Psi\rangle$ of the state $|\Psi\rangle$? This number - the multiplicity of the given occurrence of $\varepsilon$ - is equal to $ \sum\limits_i(R_{i,j}+I_{i,j}).$ These numbers can be different for an arbitrary operator $A$ and the state $|\Psi\rangle$. However, since $| \Psi\rangle$ is equilibrium with respect to $A$, $ \sum\limits_i(R_{i,j}+I_{i, j})$ will be the same for different $j$.

We introduce the notation $ \nu=\sum\limits_i(R_{i,j}+I_{i, j})$ - this is the number of occurrences of $\varepsilon$ in any column from the expansion of the matrix \eqref{hamapp}; this number $\nu$ has the order $1/\varepsilon$ at $\varepsilon\rightarrow 0$.

\bigskip

Denote by $Z_{i,j}$ the set of occurrences of the letter $\varepsilon$ in the right part of the expression \eqref{hamapp}, and let $Z_j=\bigcup_iZ_{i, j}$. Then the number of elements in the set $Z_j$ will be equal to $ \nu$.

Consider the smaller value of the amplitude quantum: $\epsilon=\varepsilon/\nu$. We substitute in the expression \eqref{quanta_expansion} instead of each occurrence of $\varepsilon$ its formal decomposition of the form $\varepsilon=\overbrace{\epsilon+\epsilon+\ldots +\epsilon}^\nu$, obtaining the decomposition of the amplitudes of the initial state into smaller numbers:

\begin{equation}
\label{refined_expansion}
\begin{array}{ll}
\la_j=&\langle j|\Psi\rangle\approx sign_{re}( \underbrace{\overbrace{\epsilon+\epsilon+\ldots +\epsilon}^\nu+\overbrace{\epsilon+\epsilon+\ldots +\epsilon}^\nu+\ldots +\overbrace{\epsilon+\epsilon+\ldots +\epsilon}^\nu}_{M_j})+\\
&sign_{im}i( \underbrace{\overbrace{\epsilon+\epsilon+\ldots +\epsilon}^\nu+\overbrace{\epsilon+\epsilon+\ldots +\epsilon}^\nu+\ldots +\overbrace{\epsilon+\epsilon+\ldots +\epsilon}^\nu}_{N_j}).
\end{array}
\end{equation}

Let $W^j_1,W^j_2,..., W^j_{M_j+N_j}$ be the sets of occurrences of the letter $\epsilon$ in the right part of the expression \eqref{refined_expansion}, marked with upper curly brackets. In each of these sets of $ \nu $ elements, as in the previously defined sets of $Z_j$. Therefore, we can construct for each such set $W^j_s$ a one-to-one mapping of the form $\xi:\ W_s^j\rightarrow Z_j$. For each occurrence of $\varepsilon$ in \eqref{quanta_expansion}, its descendants are naturally defined - the occurrences of $\epsilon$ in \eqref{refined_expansion}; the total number of descendants for each occurrence will be $\nu$.

We define the quantization of the amplitudes $\theta=\theta (\epsilon)$ so that the $ id $ of the  amplitude quanta $\kappa\in\theta$ will simply be the occurrences of $ \epsilon$ in the expansions of \eqref{refined_expansion} for all $j$. Define, as required in \eqref{quanta}, the initial state and initial type of this quantum as the state and type of this occurrence. It remains to determine the transitions of states and types. This definition is given in the following natural way.

\begin{figure}
\centering
\includegraphics[height=0.5\textwidth]{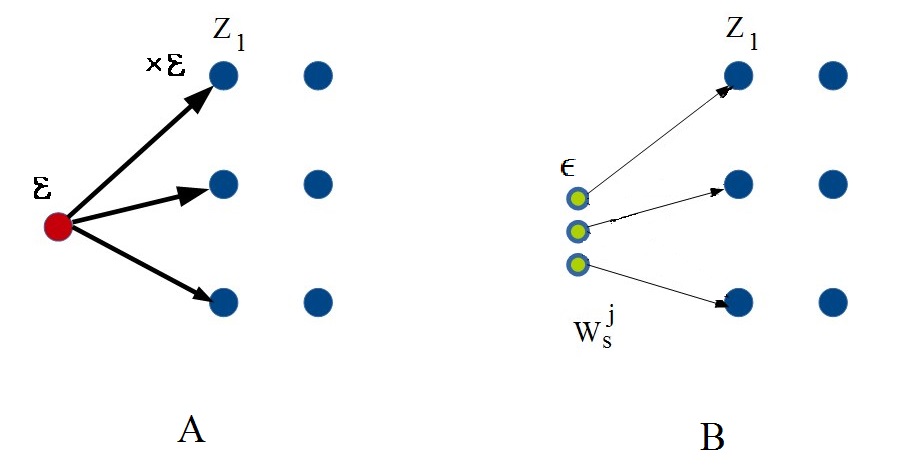}
\caption{A. Multiplying the state vector by the matrix. The contribution of each occurrence of $\varepsilon$ is multiplied by $\varepsilon$.
B. $\theta$- initial state shift. The size of the $\epsilon $ amplitude quantum is of the order of
 $\varepsilon^2$. }
\label{fig:nps}
\end{figure}

Each pair of the form $(w_s^j,\xi(w_s^j))$, where $w_s^j\in W_s^j$, will correspond to the transition of states and the transition of types in a natural way. Namely, the state transition will have the form $j\rightarrow i$ for such $i$ that $\xi(w_s^j)\in Z_{i,j}$; the transition of the types $t_{in}\rightarrow t_{fin}$ is defined so that $t_{in}$ is the type of occurrence\footnote{The type of occurrence is determined naturally after opening the parentheses, for example, for the occurrence of $...-i\epsilon ...$ the type is $-i$.} $w_s^j$, and the type $t_{fin}$ is the product of the occurrence type $t_{in}$ by the occurrence type $\xi(w_s^j)$. The sets $W^j_s$ do not intersect at different pairs $j,s$, so we consider all occurrences of the letter $\epsilon$ in the right part of\eqref{refined_expansion} to be the domain of the function definition $\xi$ (see Fig. \ref{fig:nps}).

Now let the transition of states and types for a given quantum $\kappa\in\theta$ correspond to the mapping $\xi$ in the sense defined above. The condition {\bf Q} will be met, since there are no reducing terms in the expression for the matrix element \eqref{hamapp}. So we defined the quantization of the amplitude.

Due to our definition of the function $ \xi$, the distribution of amplitudes in the state $|\theta\Psi\rangle$ will be approximately proportional to the distribution of amplitudes in the state $A|\Psi\rangle$, and the accuracy will increase indefinitely with the decrease of $\varepsilon$ to zero. In order to determine the value of the function $c (\epsilon)$ necessary for consistency of $\theta$ with the operator $A$, we calculate the contribution of each occurrence of $\varepsilon^2$ to the right side of the equality \eqref{mult} and compare it with the contribution of the corresponding letter $\epsilon$ to $ |\theta\Psi\rangle$.

\begin{figure}
\centering
\includegraphics[height=0.5\textwidth]{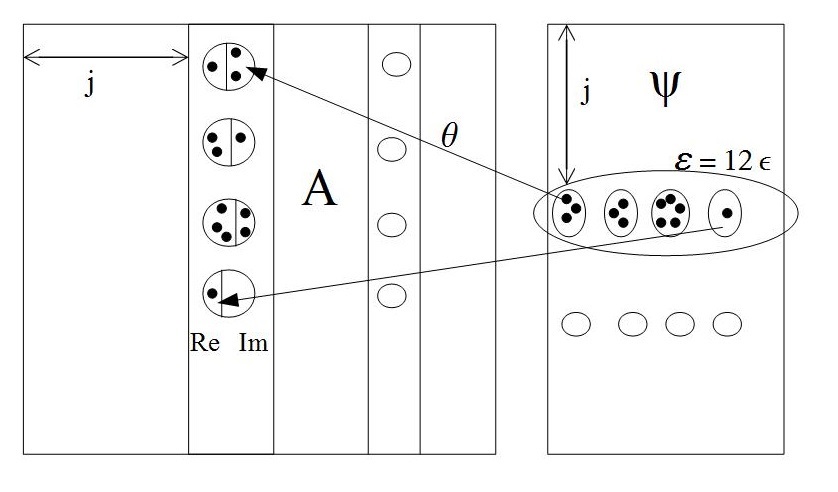}
\caption{Determinism of trajectories when multiplying the state vector $|\Psi\rangle$ by the matrix $A$. Each quantum of the ampitude passes into a certain quantum of the amplitude of the resulting state, there are no branches. }
\label{fig:matrix}
\end{figure}

Let's fix any transition of the types $t_{in}\rightarrow t_{fin}$ and the transition of the states $s_{in}\rightarrow s_{fin}$. We will call the occurrence of $\varepsilon^2$ in the result of opening brackets in \eqref{mult} corresponding to these transitions if $j=s_{in},\ i=s_{fin}$, and this occurrence is obtained by multiplying the occurrence of $\varepsilon$ of type $t_{in}$ in the first factor of the right side of \eqref{mult} by the occurrence of $\varepsilon$ in the second factor of type $t'$, so that $t_{in}t'=t_{fin}$. For each such occurrence, $\varepsilon^2 $ corresponds to exactly one quantum of the amplitude of the size $ \epsilon$ from the quantization of the amplitude defined above via the function $\xi$, which has the same transitions of states and types: this quantum corresponds to the occurrence of $\epsilon$, which is translated by a one-to-one mapping $\xi$ to this occurrence $\varepsilon^2$ (see Fig. \ref{fig:matrix}).

So, the occurrences of $\varepsilon^2$ in \eqref{mult} are in one-to-one correspondence with the occurrences of $\epsilon$ in \eqref{refined_expansion} and we obtain $c(\varepsilon)=\varepsilon\nu$. 

Note that if $A=0$, you can take $c (\varepsilon)=0$ and any quantization of the amplitude will be suitable. 

The theorem is proved.

Note that if we abandon the condition {\bf Q} and the condition of the equilibrium state $|\Psi\rangle$ with respect to $A$, we can also define matrix determinism, only we need to introduce the reducing terms $\varepsilon - \varepsilon$ into the matrix elements; then the formal entries of the amplitudes for all columns of $A$ will contain the same number of terms, and the reasoning will be valid, but interference will now occur not only between the descendants of different base states, as for equilibrium states. $|\Psi\rangle$, and also between descendants of the same state.

Note also that a quantum computation in which only gates of the form $CNOT$ and Hadamard are used has the property that for any gate the number of elements in all $Z_j$ will be the same, so that for such calculations determinism is provided with interference only between images of different basis states. This is, in particular, the Grover GSA algorithm.

Moreover, for promising implementations of quantum gates on photons (see, for example, \cite{A}), the states resulting from the implementation of gates are connected, so that interference in the course of such quantum calculations also has the property noted above.

\section{Conclusion}

We have argued for limiting the standard matrix formalism of quantum theory in the field of complex systems, as the uncertainty relation ''accuracy-complexity'' of a wave function, where the coefficient is an upper bound on the possible number of qubits that can have an irreducible entangled state. This approach does not contradict any experiments with multi-bit systems, but allows to build models of such systems on existing supercomputers. This restriction also applies to the equivalence of bases in the state space, and implies the existence of a minimum nonzero amplitude in the superposition. The size of this quantum of amplitude can be approximately found in experiments on the implementation of Grover's algorithm. The quantization of the amplitude makes it possible to introduce some type of determinism into quantum theory, which is not reduced to a quasi-classical approximation.

\section{Acknowledgements}

The paper was prepared in the Moscow Center for Fundamental and Applied Mathematics.
I am grateful to Dr. Hai Wang for the useful comments, which helped me to improve the text.

\end{document}